# Statistics of geometric clusters in the Ising model on a Bethe lattice: statistical mechanics approach

P. N. Timonin

*Physics Research Institute, Southern Federal University, 194 Stachki Avenue, Rostov-on-Don, 344090 Russia,* e-mail: pntim@live.ru

*The statistical mechanics method is developed for determination of generating function of like-sign spin clusters' size distribution in Ising model as modification of Ising-Potts model by K. K. Murata (1979). It is applied to the ferromagnetic Ising model on Bethe lattice. The analytical results for the field-temperature percolation phase diagram of + spin clusters and their size distribution are obtained. The last appears to be proportional to that of the classical non-correlated bond percolation with the bond probability, which depends on temperature and Ising model parameters.*

## I. Introduction

The properties of the like-sign (geometric) spin clusters in ferromagnetic Ising models attracted much attention very early [1-4]. The natural question has aroused – does the ferromagnetic transition results from the appearance of infinite percolation cluster of the like-sign spins? Very soon, it was found that such percolation transition may not coincide with ferromagnetic one [1, 2] and this stimulated further studies of such correlated percolation.

The basic approach to study the geometric Ising clusters was advanced by Murata [3]. He showed how the generating function of their size distribution could be obtained from the addition of special Ising - $q$-state Potts interaction with coupling $\tilde{J}$ to original Ising ferromagnetic Hamiltonian with coupling $J$. The method consists in calculating the partition function of such model in the $\tilde{J} \to \infty$ and $q \to 1$ limits. Then Coniglio and Klein [4] noticed that at $\tilde{J} = J$ Murata's model on arbitrary lattice has percolation transition right at the ferromagnetic critical point and it has the same Ising scaling indexes. But in this case, the so-called Fortuin-Kasteleyn (FK) clusters percolate. FK clusters [5] are constructed as subsets of geometric clusters via random placement of bonds in them with the temperature-dependent probability $p = 1 - e^{-J/T}$. This discovery has very strong impact leading to the new efficient Monte-Carlo algorithm for Ising or Potts models' simulations via FK clusters' update [6].

Since then the critical properties of the $\tilde{J} = J$ version of Murata's model and its generalizations has been thoroughly investigated on 2D lattices [7, 8]. Geometric clusters are also studied using Monte-Carlo simulations in 2D and 3D Euclidean lattices [9-13]. Their distribution is, in principle, observable quantity and its knowledge can be important for transport processes in magnetic materials, particularly for spintronic devices [14]. Also, percolation of simple geometric Ising clusters is very important from the theoretical point of view providing the natural example of correlated percolation [15]. Meanwhile, to our knowledge the original Murata's method has never been applied to find the size distribution of geometric clusters in any Ising model.

Here we determine the size distribution of geometric Ising clusters for ferromagnetic Ising model on Bethe lattice using the modified Murata's method, which needs only $q \to 1$ limit. Therefor its implementation can essentially simplify the numerics as the achievement of $\tilde{J} \to \infty$ limit is very time-consuming procedure. In section II, the general method is described for the ferromagnetic Ising model on arbitrary graph, in Section III the results of its implementation on the Bethe lattice ferromagnetic Ising model are presented and Section IV is devoted to the discussion and conclusions.

## II. General method

Consider a ferromagnetic Ising model on a graph with *N* sites connected by the set of edges *E*. It has usual Gibbs probability distribution function depending on the set of *N* spins **s**

$$\rho(\mathbf{s}) = e^{-\beta\mathcal{H}(\mathbf{s})} .$$

To count the clusters in spin configuration **s** we place at each edge the factor $g(s_i,\sigma_i,s_j,\sigma_j)$ depending on the Ising spins and Potts variables $\sigma = 0,1,\ldots,q-1$ at the sites that edge connect

$$g(s_i,\sigma_i,s_j,\sigma_j) = 1 + \frac{1}{4}(1+s_i)(1+s_j)\left[\delta(\sigma_i,\sigma_j) - 1\right] . \qquad (1)$$

For $s_i = s_j = 1$ this factor is $\delta(\sigma_i,\sigma_j)$ otherwise it is 1. So all sites in each cluster of positive spins have the same $\sigma$ value. Also, to count such clusters we put the factor $\zeta^{1-\delta(\sigma_k,0)}$ to all sites. This defines the following Ising-Potts partition function

$$Z_q(\zeta) = Tr_{\mathbf{s}}\rho(\mathbf{s}) Tr_{\sigma} \prod_{\langle i,j \in E \rangle} g(s_i,\sigma_i,s_j,\sigma_j) \prod_{k=1}^{N} \zeta^{1-\delta(\sigma_k,0)} . \qquad (2)$$

Using its cluster representation [4] and the equality of $\sigma$ in +spin clusters we have

$$Z_q(\zeta) = Tr_{\mathbf{s}}\rho(\mathbf{s}) \prod_{cl(\mathbf{s})} \left[1 + (q-1)\zeta^{n_{cl}(\mathbf{s})}\right] = \left\langle \left[1+(q-1)\zeta\right]^{N_-(\mathbf{s})} \prod_{n=1}^{\infty} \left[1+(q-1)\zeta^n\right]^{N_{+,n}(\mathbf{s})} \right\rangle Z_1$$

Here $n_{cl}(\mathbf{s})$ denotes the capacity of a cluster that is present in configuration **s**, $N_{+,n}(\mathbf{s})$ is the number of *n*-site +spin clusters in the configuration, $N_-(\mathbf{s})$ is the number of negative spins in it and

$$Z_1 = Tr_{\mathbf{s}} e^{-\beta\mathcal{H}(\mathbf{s})}, \quad \langle A(\mathbf{s}) \rangle \equiv Tr_{\mathbf{s}}\rho(\mathbf{s}) A(\mathbf{s}) / Tr_{\mathbf{s}'} e^{-\beta\mathcal{H}(\mathbf{s}')} .$$

Hence, we have for $q \to 1$

$$\frac{1}{N}\ln\frac{Z_q(\zeta)}{Z_1} \approx (q-1)\left[G_+(\zeta) + \zeta\left\langle\frac{N_-(\mathbf{s})}{N}\right\rangle\right] = (q-1)\left[G_+(\zeta) + \frac{\zeta}{2}(1-m)\right], \qquad (3)$$

where *m* denotes the average magnetization

$$m = \frac{1}{N}\sum_{i=1}^{N}\langle s_i \rangle = \frac{1}{N}\frac{d\ln Z_1}{dh} \qquad (4)$$

and

$$G_+(\zeta) = \sum_{n=1}^{\infty} \zeta^n v_n^+, \quad v_n^+ = \left\langle\frac{N_{+,n}(\mathbf{s})}{N}\right\rangle$$

is the generating function for the numbers (per site) of *n*-site +spin clusters $v_n^+$.

## III. Application to the Ising model on a Bethe lattice

We consider Ising model on Bethe lattice with the nearest-neighbor ferromagnetic interactions. Its Gibbs function is

$$\rho(\mathbf{s}) = \exp\left( K \sum_{\langle i,j \rangle} s_i s_j + h \sum_i s_i \right), \quad K = J/T > 0, \quad h = H/T$$

*J* is the exchange constant, *H* – magnetic field and *T* – temperature.

To find $Z_q(\zeta)$ for it we introduce the partial partition function $V_l(s,\sigma)$ for the *l*-th generation Caley tree summed over all Ising and Potts variables except the root ones. They obey the following recursion relations

$$V_{l+1}(s,\sigma) = \sum_{s'=\pm 1} \sum_{\sigma'=0}^{q-1} B(s,\sigma,s',\sigma') A(s',\sigma') V_l^{z-1}(s',\sigma'), \qquad (5)$$

where bond and site operators for our model are

$$B(s,\sigma,s',\sigma') = e^{Kss'} g(s,\sigma,s',\sigma'), \quad A(s,\sigma) = e^{hs} \zeta^{1-\delta(\sigma,0)}$$

so we have

$$V_{l+1}(s,\sigma) = \sum_{s'=\pm 1} e^{(Ks+h)s'} \sum_{\sigma'=0}^{q-1} g(s,\sigma,s',\sigma') \zeta^{1-\delta(\sigma',0)} V_l^{z-1}(s',\sigma'). \qquad (6)$$

The solution to this equation admits the following representation

$$V_l(s,\sigma) = e^{a_l + f_l s} \mu_l^{\delta(s,1)[1-\delta(\sigma,0)]} \qquad (7)$$

From (6) and (7) we get

$$f_{l+1} = \frac{1}{2} \ln \frac{e^{-(z-1)f_l - h - K} \left[1 + (q-1)\zeta\right] + e^{(z-1)f_l + h + K}}{e^{-(z-1)f_l - h + K} \left[1 + (q-1)\zeta\right] + e^{(z-1)f_l + h - K} \left[1 + (q-1)\zeta \mu_l^{z-1}\right]} \qquad (8)$$

$$\mu_{l+1} = \frac{e^{-(z-1)f_l - h - K} \left[1 + (q-1)\zeta\right] + e^{(z-1)f_l + h + K} \zeta \mu_l^{z-1}}{e^{-(z-1)f_l - h - K} \left[1 + (q-1)\zeta\right] + e^{(z-1)f_l + h + K}} \qquad (9)$$

$$a_{l+1} = (z-1)a_l + \frac{1}{2} \ln\left[ e^{-(z-1)f_l - h - K} \left[1 + (q-1)\zeta\right] + e^{(z-1)f_l + h + K} \right] +$$
$$+ \frac{1}{2} \ln\left[ e^{-(z-1)f_l - h + K} \left[1 + (q-1)\zeta\right] + e^{(z-1)f_l + h - K} \left[1 + (q-1)\zeta \mu_l^{z-1}\right] \right] \qquad (10)$$
$$\frac{1}{2} \ln\left[ 2\cosh\left[(z-1)f_l + h + K\right] + (q-1)\zeta e^{-(z-1)f_l - h - K} \right] +$$
$$+ \frac{1}{2} \ln\left[ 2\cosh\left[(z-1)f_l + h - K\right] + (q-1)\zeta \left[ e^{-(z-1)f_l - h + K} + e^{(z-1)f_l + h - K} \mu_l^{z-1} \right] \right]$$

At $l \to \infty$ $f_l$, $\mu_l$ tend to the stable points. Denoting them in $q \to 1$ limit as

$$\lim_{l \to \infty} f_l \approx f + (q-1)g, \quad \lim_{l \to \infty} \mu_l = \mu + O(q-1), \qquad (11)$$

we get for *f* from Eq. (8) the usual Ising equation of state

$$f = \frac{1}{2}\ln\frac{\cosh[(z-1)f+h+K]}{\cosh[(z-1)f+h-K]} = \tanh^{-1}\{\tanh K \tanh[(z-1)f+h]\}. \tag{12}$$

The solutions to this equation must obey the stability condition

$$\tanh^2 f > \tanh K \tanh(K-K_c), . \tag{13}$$

$$K_c = J/T_c = \frac{1}{2}\ln\frac{z}{z-2}$$

Here $T_c$ is the temperature of ferromagnetic transition.

Percolation equation of state for $\mu$ follows from Eq. (9)

$$\mu = 1 - p(K) + \zeta p(K)\mu^{z-1}, \tag{14}$$

$$p(K) = \left[e^{-2[(z-1)f+h+K]}+1\right]^{-1} = \frac{1}{2}\{1+\tanh[(z-1)f+h+K]\}$$

The stability condition for this equation is

$$(z-1)\zeta p(K)\mu^{z-2} < 1 \tag{15}$$

For *g* in Eq. (11) we get from Eq. (8)

$$g = \frac{\zeta}{2}\frac{p(-K)-p(K)-p(-K)\mu^{z-1}}{1+(z-1)[p(-K)-p(K)]} \tag{16}$$

It follows from Eq. (10)

$$\lim_{l\to\infty}[a_{l+1}-(z-1)a_l] \approx b+(q-1)c \tag{17}$$

$$b = \frac{1}{2}\ln 4\cosh[(z-1)f+h+K]\cosh[(z-1)f+h-K] \tag{18}$$

$$c = \frac{\zeta}{2}[2-p(K)-p(-K)+p(-K)\mu^{z-1}]+(z-1)g[p(K)+p(-K)-1] \tag{19}$$

Now we have all that is needed to find the free energy density for our model.

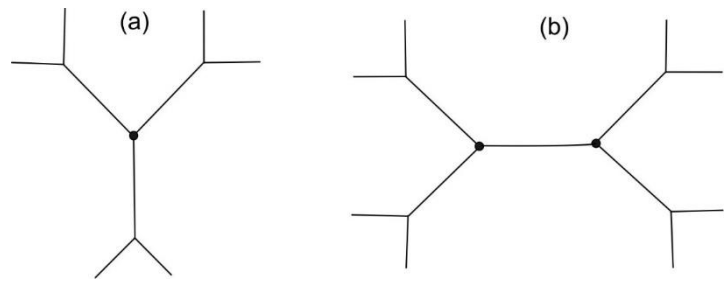

Fig. 1. Two ways of Bethe lattices construction form Caley trees, (a) to add one site to *z* trees, (b) to add two sites and a bond to 2(z-1) trees.

For the sake of self-consistency, we briefly describe the procedure of finding the free energy density for general spin model on Bethe lattice [16, 17]. Let us take $2z(z-1)$ Caley trees of *l*-th generation. There are two ways to construct Bethe lattices from them. First, adding one site to each of *z* trees, see Fig. 1a, we get $2(z-1)$ lattices with the free energy

$$F_{1,l} = 2(z-1)\ln Tr_{s,\sigma} A(s,\sigma) V_l^z(s,\sigma).$$

Second, adding two sites and a bond to each $2(z-1)$ trees, see Fig. 1b, we get *z* lattices with the free energy

$$F_{2,l} = z \ln Tr_{s,\sigma,s',\sigma'} V_l^{z-1}(s,\sigma) A(s,\sigma) B(s,\sigma,s',\sigma') A(s',\sigma') V_l^{z-1}(s',\sigma') = z \ln Tr_{s,\sigma} V_l^{z-1}(s,\sigma) A(s,\sigma) V_{l+1}(s,\sigma)$$

The last equation here follows from Eq. 5 for $V_{l+1}$. These two constructions provides two sets of the similar Bethe lattices that differ only by the number of sites, in the first case we added $2(z-1)$ sites while in the second case $2z$ sites are added (that is, two sites more). Hence, for the free energy density $N^{-1}\ln Z$ we have in the thermodynamic limit

$$2\lim_{N\to\infty} N^{-1}\ln Z = \lim_{l\to\infty}\left[F_{2,l} - F_{1,l}\right] = \lim_{l\to\infty}\left\{z\ln Tr_{s,\sigma} V_l^{z-1}(s,\sigma) A(s,\sigma) V_{l+1}(s,\sigma) - 2(z-1)\ln Tr_{s,\sigma} A(s,\sigma) V_l^z(s,\sigma)\right\}$$

For our model we have

$$\lim_{N\to\infty}\frac{2}{N}\ln Z_q(\zeta) = z\lim_{l\to\infty}\left[a_{l+1} - (z-1)a_l\right] + (2-z)\lim_{l\to\infty}\ln\sum_s e^{(zf_l+h)s}\sum_\sigma \zeta^{[1-\delta(\sigma,0)]}\mu_l^{z\delta(s,1)[1-\delta(\sigma,0)]}$$

For *q* close to unity we obtain

$$\lim_{N\to\infty}\frac{2}{N}\ln Z_q(\zeta) \approx zb + (2-z)\ln 2\cosh(zf+h) + (q-1)\left\{zc + (2-z)\left[zmg + \frac{\zeta}{2}(1-m) + \frac{\zeta}{2}(1+m)\mu^z\right]\right\} \quad (20)$$

$$\lim_{N\to\infty}\frac{2}{N}\ln Z_1(\zeta) = zb + (2-z)\ln 2\cosh(zf+h), \quad (21)$$

$$m = \lim_{N\to\infty}\frac{1}{N}\frac{d\ln Z_1}{dh} = \tanh(zf+h) \quad (22)$$

To derive the last equation for *m* we used the relation, which follows from the Ising equation of state (12)

$$\tanh(zf+h) = \frac{\tanh\left[(z-1)f+h+K\right] + \tanh\left[(z-1)f+h-K\right]}{2 - \tanh\left[(z-1)f+h+K\right] + \tanh\left[(z-1)f+h-K\right]} = \frac{p(K)+p(-K)-1}{1+p(-K)-p(K)} \quad (23)$$

Thus from (3) and (16-22) we get the generating function for the size distribution of the + spin clusters in terms of two solutions to Ising (*f*) and percolation ($\mu$) equations of state (12, 14)

$$G_+ = \frac{\zeta}{4}\mu^{z-1}\left[zp(-K)(1-m) + (2-z)(1+m)\mu\right] \quad (24)$$

This expression can be simplified via the introduction of new variable *u* instead of $\mu$

$$u = \mu + p(K) - 1 \quad (25)$$

Then the percolation equation of state (14) now reads

$$u = \zeta p(K)\left[u + 1 - p(K)\right]^{z-1} \quad (26)$$

and the stability condition for it is

$$(z-2)u < 1 - p(K) \qquad (27)$$

Then using (23) we get $G_+(\zeta)$ as the simple quadratic polynomial in $u(\zeta)$

$$G_+[u(\zeta)] = \frac{1+m}{4p(K)} u[2 - 2p(K) - (z-2)u] \qquad (28)$$

Eqs. (26-28) describe the percolation properties of + spin clusters. Note that they coincide with those for classical uncorrelated bond percolation [19] with bond probability $p(K)$ except for prefactor $(1+m)/2$ in Eq. (28). This proportionality of G to the classical one was first found in [2].

Thus, we can find the number of + spin clusters (per site)

$$N_{cl}^+ = \sum_{n=1}^{N} v_n^+ = G_+[u(\zeta=1)] = \frac{1+m}{4p(K)} u_1[2 - 2p(K) - (z-2)u_1]. \qquad (29)$$

Here $u_1 \equiv u(\zeta=1)$ obeys the equation, which follows from (26)

$$u_1 = p(K)[u_1 + 1 - p(K)]^{z-1}. \qquad (30)$$

The number of sites belonging to the finite + spin clusters is

$$N_{+sites} = \sum_{n=1}^{N} n v_n^+ = \partial_\zeta G[u(\zeta)]\big|_{\zeta=1} = \partial_u G_+[u(\zeta)] \partial_\zeta u\big|_{\zeta=1} = \frac{1+m}{2p(K)} u_1[u_1 + 1 - p(K)], \qquad (31)$$

and the density of the infinite cluster of + spins

$$P_+ = \frac{1+m}{2} - N_{+sites} = \frac{1+m}{2p(K)}(u_1 + 1)[p(K) - u_1]. \qquad (32)$$

as $\frac{1+m}{2}$ is the average density of all sites with + spins.

There is always the solution $u_1 = p(K)$ to Eq. (30) that is stable when

$$(z-1)p(K) < 1. \qquad (33)$$

It corresponds to the non-percolating phase as according to (31) $N_{+sites} = \frac{1+m}{2}$ so $P_+ = 0$ in it. Also the density of clusters (29) in this phase is

$$N_{cl}^+ = \frac{1+m}{4}[2 - zp(K)]. \qquad (34)$$

The stability condition (32) can be represented as

$$(z-2)e^{2[(z-1)f + h + K]} < 1 \qquad (35)$$

implying $f < 0$ according to (12). It follows from Eqs. (12, 22) that $sign(m) = sign(f)$, which means that percolating transition appears only in two thermodynamic phases with $m < 0$ (stable at H < 0 and metastable at $H > 0$). Meanwhile in $m > 0$ phases the solution $u_1 = p(K)$ cannot be stable so there is always the giant cluster of + spins.

It follows from (35) that in $m < 0$ phases the percolating cluster of + spins emerges when

$$f = \frac{-2(h+K) - \ln(z-2)}{2(z-1)}$$

Substituting this $f$ into Ising equation of state (12), we get the critical field $H_p(z,T)$ for percolation transition

$$H_p(z,T) = (z-2)J + \frac{T}{2}\left[(z-1)\ln\left(\frac{1+(z-2)e^{-\frac{2J}{T}}}{z-1}\right) - \ln(z-2)\right]. \qquad (36)$$

The phases with $m < 0$ are stable at (cf. Eq. (13))

$$\tanh f < \tanh f_- = -\sqrt{\tanh K \tanh(K - K_c)}\;.$$

The field $H(z, f_-, T)$ found from the equation of state (12) defines the field $H_-(z,T) = \max\left[H(z, f_-, T), 0\right]$ below which phases with $m < 0$ are stable. Numerics shows that $H_p(3,T) > H_-(3,T)$ so at z = 3 the $m < 0$ phases are always in the non-percolating state, while at z > 3 $H_p(z,T) < H_-(z,T)$ for all T and percolating transition takes place, see Fig. 2.

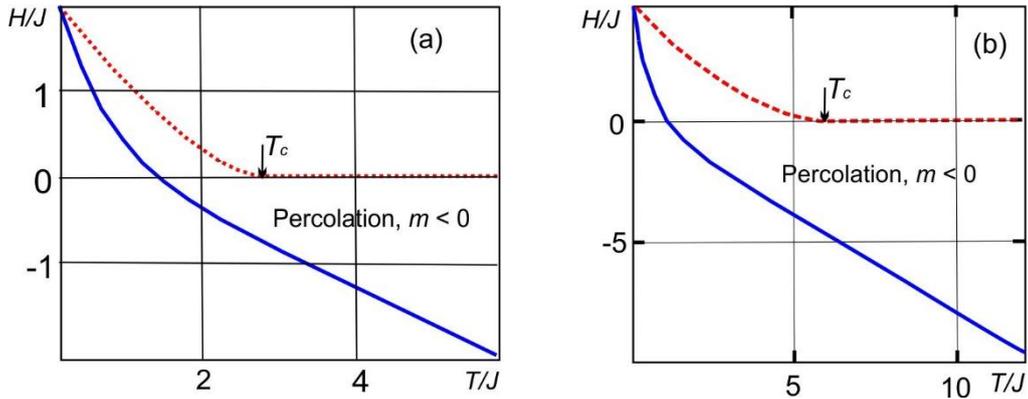

Fig. 2. Percolation phase diagrams for + spin clusters on Bethe lattices with z=4 (a) and z=7 (b). Above solid lines the giant cluster of + spins emerges. Dashed lines are the stability limits of $m < 0$ phases in which this cluster exists.

At last, $G_+$ (28) allows to obtain the explicit expression for $v_n^+$ in terms of one Ising thermodynamic parameter. This can be made with the Lagrange inversion formula for expansion of implicitly defined function [18]. Actually, it amounts to the change of variable and integration by parts in the standard expression for $v_n^+$. We have

$$v_n^+ = \oint \frac{d\zeta}{2\pi i} \zeta^{-n-1} G_+[u(\zeta)] = \oint \frac{du}{2\pi i} \frac{\partial \zeta}{\partial u} \zeta(u)^{-n-1} G_+(u) = \frac{1}{n}\oint \frac{du}{2\pi i} \zeta(u)^{-n} \partial_u G_+(u) =$$

$$= \frac{(1+m)}{2} \frac{p^{n-1}(K)}{n} \oint \frac{du}{2\pi i} u^{-n}[u+1-p(K)]^{n(z-1)}[1-p(K)-(z-2)u]$$

The last integration here is easily performed with the result

$$v_n^+ = \frac{1+m}{2} \frac{z}{n[n(z-2)+2]} \binom{n(z-1)}{n-1} p^{n-1}(K)[1-p(K)]^{n(z-2)+2}. \qquad (37)$$

Except for the factor $\frac{1+m}{2}$ this expression coincides with that of the classical non-correlated bond percolation with the bond probability $p(K)$ which has been derived by the different method in [19]. Accordingly, the + clusters percolation has the same classical critical indexes. Note also that $G_-(\zeta, K, h) = G_+(\zeta, K, -h)$ describes the minus-spin percolation as Ising Hamiltonian is invariant under changes $H \to -H$, $\mathbf{s} \to -\mathbf{s}$. Accordingly, the phase diagram for the minus-spin percolation is obtained from Fig. 2 by reflecting it with respect to horizontal axis $H = 0$, while $v_n^-$ for minus-spin clusters we get from (37) substituting $p(K, f, h) \to p(K, -f, -h)$, $\frac{1+m}{2} \to \frac{1-m}{2}$.

### IV. Discussion

We show that the calculation of specific Ising-Potts partition function is the useful method to study the percolation of geometric Ising clusters. With it the percolation phase diagram and size distribution of + spin clusters in ferromagnetic Ising model on Bethe lattice are found analytically. The size distribution in such correlated percolation appears to be proportional to that of the classical non-correlated bond percolation. It seems that this result is not solely the property of the Bethe lattice model as the last is the good mean-field approximation for graphs and lattices with $z \gg 1$ outside the critical region. Probably, in the mean-field percolation region of a wide class of the Ising ferromagnetic models the correlations amount to the formation of independent pairs of nearest neighbor + spins.

Comparing with the existing results for Ising clusters percolation on Bethe lattice [2, 3] we should note that $v_n^+$ found in [2] have prefactor different from $\frac{1+m}{2}$ in (37). Accordingly, our result for the cluster's density (29) also differs by a prefactor from that in [2]. In [3] $v_n^+$ were not determined, only global characteristics were obtained, i.e. $N_{cl}^+, N_{+sites}^+, P_+$. They are explicitly the same as our results (29, 31, 32) as well as the equations of state (12, 26) (note that in this paper $x = \langle n \rangle = \frac{1+m}{2}$). In both papers the critical field (36) and phase diagram (Fig. 2) were not determined while this can be easily done using the equations of state (12, 26).

The present approach can be useful for the numerical studies of Ising clusters' percolation on Euclidean lattices being more preferable than that of [3] as it does not need the fulfilment of infinite coupling limit. Thus, for rough estimate of Ising clusters' size generation function the usual Monte Carlo simulations can be used to obtain $Z_q(\zeta)$ for several integer $q$ and to interpolate it to $q = 1$. To get more precise results one should extend the expression (2) for $Z_q(\zeta)$ to real $q$. This can be done, for example, within the transfer matrix representation of $Z_q(\zeta)$, see Refs. [20, 21]. Note also, that present method can be easily modified for other types of Ising clusters.

### Acknowledgment

This work was done under financial support by the Ministry of Education and Science of the Russian Federation (state assignment Grant No. 3.5710.2017/8.9).